\documentclass{WileyMSP-template}
\usepackage{xcolor}

\usepackage{gensymb}
\usepackage{textcomp}
\begin{document}

\pagestyle{fancy}

\rhead{\includegraphics[width=2.5cm]{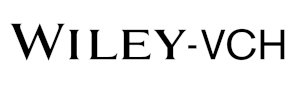}}

\title{Blue-band frequency comb and photodarkening in silica whispering gallery microresonators}


\maketitle


\author{Ke Tian,}
\author{Jibo Yu,}
\author{Fuchuan Lei,}
\author{Jonathan M. Ward,}
\author{Angzhen Li,}
\author{Pengfei Wang, and}
\author{S\'ile {Nic Chormaic*}}


\begin{affiliations}
K. Tian, J. Yu, A. Li, S. {Nic Chormaic}\\
Okinawa Institute of Science and Technology Graduate University, Onna, Okinawa 904-0495, Japan\\
E-mail: sile.nicchormaic@oist.jp

F. Lei\\
Department of Microtechnology and Nanoscience, Chalmers University of Technology, SE-41296 Gothenburg, Sweden

J. M. Ward\\
Physics Department, University College Cork, Cork, Ireland

K. Tian, J. Yu, A. Li, P. Wang\\
Key Laboratory of In-Fiber Integrated Optics of Ministry of Education, College of Physics and Optoelectronic Engineering, Harbin Engineering University, Harbin 150001, China

P. Wang\\
\justifying \noindent Key Laboratory of Optoelectronic Devices and Systems of Ministry of Education and Guangdong Province College of Optoelectronic Engineering, Shenzhen University, Shenzhen 518060, China

\end{affiliations}



\begin{abstract}

\justifying \noindent To date there are extensive studies of optical nonlinearities in whispering gallery resonators (WGRs) in the near and mid-infrared wavelengths. Pushing this research into the visible region is equally valuable. Here, we demonstrate a Kerr frequency comb and Raman lasing at 462~nm in an SiO$_2$ WGR. Notably, due to the high optical intensities achieved, photodarkening is unavoidable and can quickly degrade the optical quality of both the coupling optical nanofiber and the microcavity even at very low pump powers. Nonetheless, stable stimulated Raman scattering (SRS) and hyper-parametric oscillation in normally dispersed WGRs is demonstrated in the presence of photodarkening by taking advantage of \textit{in-situ} thermal bleaching. These observations highlight the challenges of silica-based, short wavelength nonlinear optics in high quality, small mode volume  devices. We propose a method to overcome this apparent limitation and demonstrate  blue-band nonlinear optical processes in silica WGRs, thus providing a baseline for optics research in the blue region for any optical devices fabricated from SiO$_2$.

\end{abstract}


\section{Introduction}
\justifying \noindent Silica (SiO$_2$) is widely used for building both passive and active photonic devices and is a material of choice by many~\cite{doi:10.1063/1.373805}. Its properties of ultralow loss and broadband transparency make it an excellent  platform for quantum optics, nonlinear optics, and micro/nanophotonics. Silica is frequently used for fabricating whispering gallery microresonators (WGRs) \cite{silica_microsphere,silica_chip}. Such devices have emerged as near ideal for applications in nonlinear optics \cite{kippenberg2004kerr, Yang:16}, cavity quantum electrodynamics \cite{aoki2006observation,lei2020polarization}, and sensing \cite{Baaske2014, Yang:2016CRUS, Hogan2019}. In addition, silica is used for the fabrication of many commercial optical fibers from which tapered optical fibers or optical nanofibers (ONF) can be fabricated \cite{Tong2003ONF}. These ONFs are used as couplers for WGRs \cite{Knight:97,lei2020enhanced} and they have also been exploited for numerous other applications such as supercontinuum generation \cite{birks2000supercontinuum} and third-order parametric down-conversion \cite{PhysRevA.101.033840}. Notably, silica is also widely used as a cladding material for the integration of optical waveguides with high refractive indices, such as silicon nitride \cite{moss2013new,morin2021cmos,ye2021overcoming} and lithium niobate  \cite{jian2018high}. 

\justifying Conventionally, most studies on silica-based photonics, especially involving fiber optics, are conducted in the infrared (IR) wavelength band. However, there is significant motivation in pushing silica photonics to the near-IR and visible wavelengths. Developing  applications in many areas such as biological imaging, underwater communication and detection, atomic clocks, and quantum technologies all rely on wavelengths in this range. To date, near-visible optically pumped parametric oscillation and Kerr frequency combs have been realized in SiO$_2$ WGRs \cite{savchenkov2011kerr,bubble_comb,lee2017towards,ShoComb2018, ma2019visible,Savchenkov:20}. Green and blue light generation through a third-harmonic process with near-infrared pumping near 1550 nm has also been observed \cite{carmon2007visible,fujii2017third,chen2020chaos}. Aside from photonics, the desire to use blue light in silica devices for quantum applications is also increasing. Recently, the remote entanglement of two strontium ions via the polarization of two spontaneously emitted photons at 422~nm using a conventional single-mode fiber link was demonstrated \cite{Stephenson20}. For neutral atoms, low power blue light at 482 nm propagating through an SiO$_2$ ONF was used to generate cold rubidium Rydberg atoms via a two-photon transition in the evanescent field region with the aim of developing an all-fibered quantum gate \cite{RajasreeRydberg}. However, studies on the effects of blue light on the properties of tapered optical fibers or on nonlinear optical processes in silica resonators is very limited, through clearly very desirable at this time.

\justifying Here, we studied nonlinear effects in a silica microsphere:nanofiber coupled system pumped in the blue wavelength region. Blue-band Raman lasing and an optical frequency comb was generated in the resonators even in the presence of normal dispersion and photodarkening. The creation of color centers from two-photon absorption in the blue-band caused excess absorption loss. As a result, the ability of the nanofiber to transmit blue light and the optical quality of the whispering gallery modes (WGMs) appeared to degrade rapidly. Crucially, the photodarkening in nanofibers was shown to have a sufficiently low temperature dependence, so that a small increase in ambient temperature led to effective thermal bleaching. This provided a means to recover the optical quality of the cavity and allowed for the generation of stable Raman lasing and an optical frequency comb.

\section{Experiment and Results}
\subsection{Photodarkening in Silica Nanofibers}
\begin{figure*} [ht]
\centering
      \includegraphics[width=1.0\textwidth]{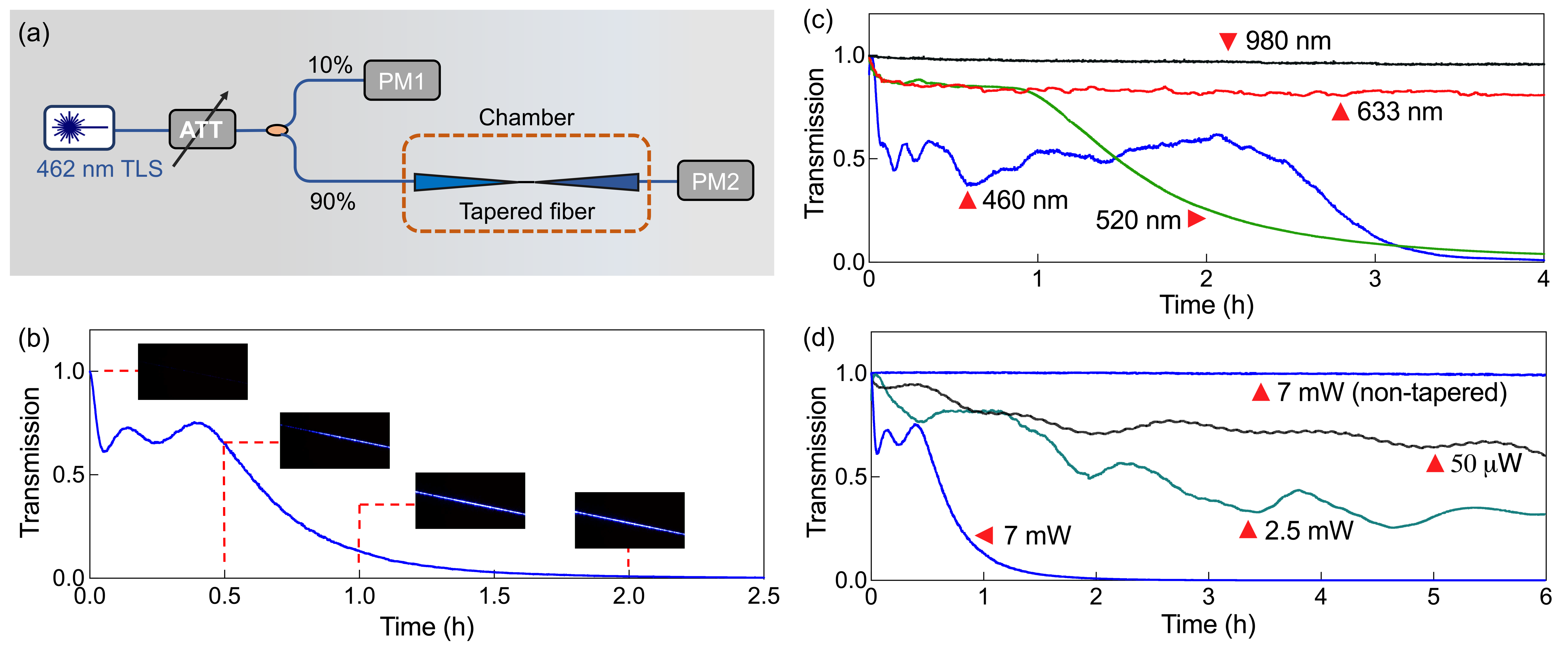}
  \caption{{Photodarkening in optical nanofibers} (a) Experimental setup for measuring the transmission through a nanofiber. TLS: tunable laser system; ATT: attenuator; PM: power meter. (b) Normalized transmission through the 460HP nanofiber (500 nm waist) for a fixed pump power of 7~mW at 462 nm. Inset: images of the waist of the nanofiber over times of 0/30/60/120 minutes, the exposure time of the CCD camera was set to the lowest value at the beginning. (c) Normalized transmission through the 980HP nanofiber (500 nm waist) for a fixed pump power of 7~mW and different wavelengths. (d) Normalized transmission through the 460HP nanofiber (500~nm waist) for a fixed wavelength of 460~nm and different pump powers. The transmission of an untapered fiber is given for comparison.
  \label{fig.1}}
\end{figure*}

First, we studied the transmission of blue light in an optical nanofiber. The experimental setup is schematically illustrated in Fig. \ref{fig.1}(a). A tunable laser (Toptica Photonics, DL Pro HP 461) with a center wavelength of 462 nm was used as the light source. The light was coupled into a single-mode fiber (SMF, 460HP, Thorlabs), then passed through an attenuator (ATT) and a 90/10 inline beam splitter (BS). The 10$\%$ port was connected to a power meter and the 90$\%$ port was connected to the nanofiber. The output power from the nanofiber was recorded by another power meter. The nanofiber, with a minimum diameter at the waist of 500 nm, was made from the same 460HP fiber using the heat-and-pull method with a ceramic microheater \cite{microheater}. The nanofiber was placed in an enclosed chamber to avoid environmental disturbances.

Figure \ref{fig.1}(b) shows the transmission through the nanofiber at a fixed pump power of 7~mW. The transmission dropped rapidly to 62\% in 3 minutes, then oscillated for 20 minutes, and finally decayed close to zero after 2 hours. A CCD camera was used to image and record the evolution of the scattered light from the nanofiber and images at different time intervals are shown as insets in Fig. \ref{fig.1}(b). As the loss of the 462 nm light in the nanofiber increased with time, the  brightness of the scattered light also gradually increased. Analysing the images reveals the number of scattering points activated over time for a length of ~900~$\mu$m along the taper waist. Initially (first image in Fig. \ref{fig.1}(b)), there are approximately ten dim scattering points on the waist of the nanofiber. As time progresses these points become brighter and new points appear, increasing to a least 100 bright scattering points after 2 hours, with about one scatterer every 9~$\mu$m. It is worth emphasizing that the observed scattered light was not caused by dust particles, since the nanofiber was kept in a clean chamber and a similar phenomenon was not observed when pumping at 980 nm. After irradiation, the nanofiber was examined with a scanning electron microscope (SEM) and no trace of contamination or damage was found on its surface. These observations lead us to believe the loss in transmission of the nanofiber was caused by photodarkening resulting from the photoactivation of color centers.


In order to better understand the photodarkening wavelength dependence in very thin tapered fibers, the transmission of light through a 500~nm waist nanofiber as a function of time was characterized  using lasers with wavelengths  of 980~nm, 633~nm, 520~nm and 462 nm. A 980HP fiber (core diameter 3.6~$\mu$m) was used since it has a larger core than the 460HP fiber (core diameter 2.5~$\mu$m) used in the previous tests. This ensured easier coupling of light into the fiber over a wide range of wavelengths. The pump power coupled into the fiber was 7~mW for each wavelength and was kept constant throughout. A typical normalized transmission signal as a function of time is shown in Fig. \ref{fig.1}(c). It can be seen that, for this particular fiber sample, the transmission  decreased at different rates for the different wavelengths. The reduction in the transmission at 980~nm was slow and  only 0.04\% over the test time (more than 4 hours). At 633 nm, the light initially decayed rapidly with an exponential time constant of ~6  minutes; however, after 20 minutes, the normalized transmission remained nearly fixed at 80\%. At shorter wavelengths the behavior was noticeably different; the transmission for 520~nm was similar to that for 633~nm for the first hour, but it decreased to 4\% after four hours with an exponential time constant of 50 minutes. For 460 nm light, the transmission dropped rapidly to 62\% in 4 minutes, then oscillated for 2.5 hours, and finally decayed to 1\% after 4 hours with an exponential time constant of 25 minutes. These results demonstrate that, for the same input power, the power drop at shorter wavelengths was much more significant and displayed unusual characteristics. This phenomenon followed the same trend as photodarkening in untapered doped fibers \cite{wavelength,thermal-bleaching2} and points to photodarkening being the dominant process behind the observed behavior in our nanofibers.

Excess optical loss caused by photodarkening in untapered fibers can be  related to the intensity of the pump light \cite{power}. Therefore, we investigated the transmission through a nanofiber at a wavelength of 462 nm over time for different input powers. The nanofibers were made from 460HP fiber and the diameter was approximately 500 nm in each case.  The results are shown in Fig. \ref{fig.1}(d) and, as a reference, the transmission through an untapered 460HP fiber is also presented. The transmission of the untapered fiber was reduced by only 0.9\% over 6 hours for an input power of 7 mW. However, for the tapered fibers, their transmission was unstable when the input power exceeded 50 $\mu$W. As the power increased from 50 $\mu$W to 7 mW, the power downtrend was more dramatic. Although the transmission can remain stable for lower powers for a reasonable period of time, 50 $\mu$W is generally not sufficient to reach the threshold needed for the observation of nonlinear effects in WGRs, such as SRS and hyper-parametric oscillation. In contrast to the studies on photodarkening in commercial optical fibers when using high powers (typically a few Watts) \cite{mode_laser,Amplifier,stone}, the threshold for photodarkening in the nanofibers reported herein was very low ($\sim$50 $\mu$W). The most likely explanation is the increased probability of two-photon absorption due to the very high optical intensities (MW/cm$^2$ - GW/cm$^2$) at the waist of the tapered nanofiber.


\subsection{Blue-band Nonlinear Optics in a Nanofiber-coupled Silica WGR }
\begin{figure*} [ht]
\centering
      \includegraphics[width=0.7\textwidth]{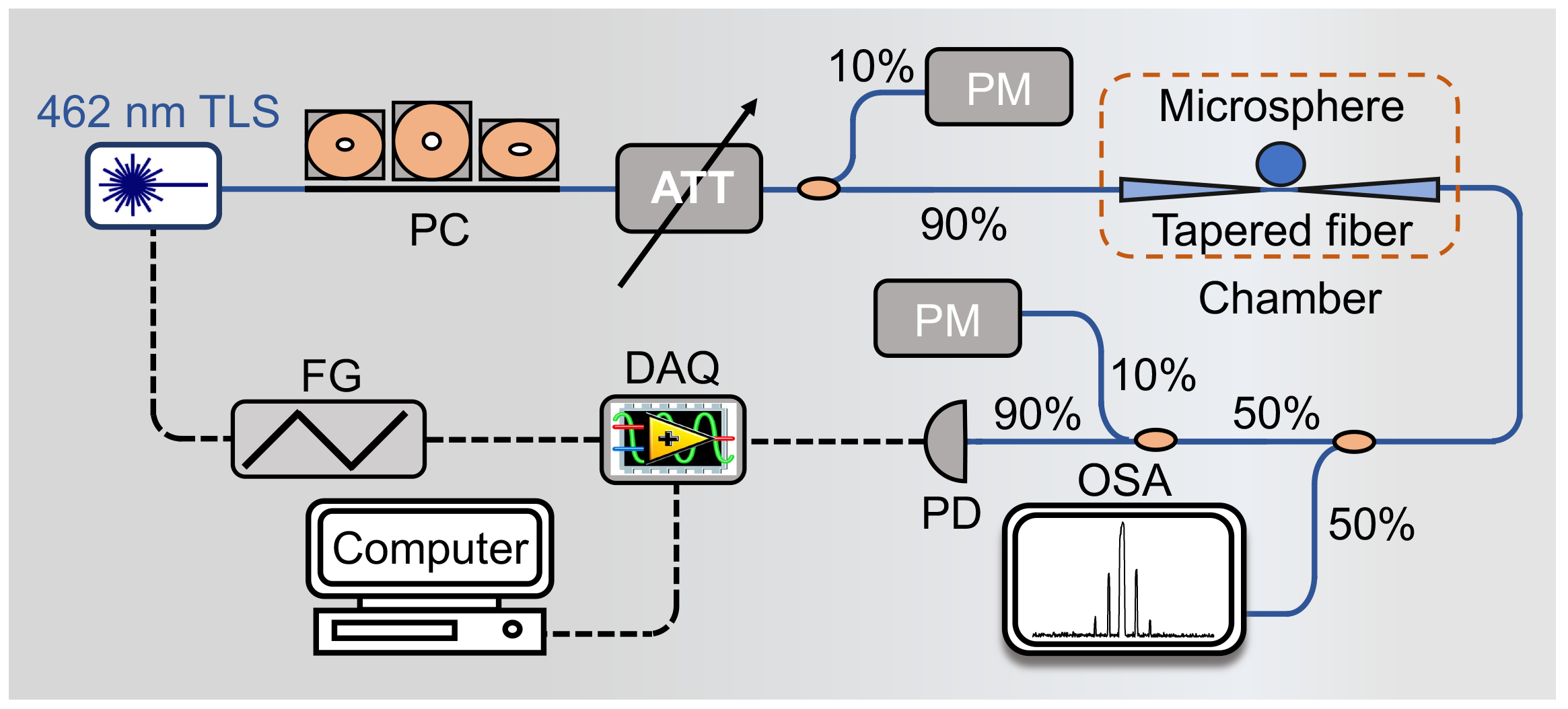}
  \caption{{Experimental setup for studying nonlinear optical effects in a nanofiber-coupled silica microsphere.} The blue solid lines represent the optical path and the black dashed lines are the electrical connections. TLS: tunable laser system; PC: polarization controller; ATT: attenuator; PM: power meter; FG: function generator; DAQ: digital acquisition card; PD: photodetector; OSA: optical spectrum analyzer.  
  \label{fig.setup}}
\end{figure*}

Tapered optical fibers are frequently used as couplers to excite modes in whispering gallery resonators. When pump light is coupled from a fiber to the resonator, the optical intensity may be further amplified in the cavity due to its small mode volume and high optical quality factor (Q-factor). To investigate the influence of photodarkening on nonlinear effects in a silica nanofiber-coupled WGR system, a straightforward SRS measurement over time was taken. The experimental setup for measuring various nonlinear effects (SRS, hyper-parametric oscillation, and frequency combs) is schematically illustrated in Fig. \ref{fig.setup}. The same 462 nm tunable laser as used in the nanofiber transmission measurement was used as pump light to excite the nonlinear optical processes in a SiO$_2$ microsphere. The light was coupled into a single-mode fiber (SMF, 460HP, Thorlabs), then passed through a polarization controller and a 90/10 inline beam splitter. As before, the 10$\%$ port was monitored on a power meter and the 90$\%$ port was spliced to the tapered fiber. The tapered nanofiber and the microsphere  were placed in an enclosed chamber to avoid environmental disturbance. Another 50/50 inline beam splitter was connected to the output of the nanofiber; one port was connected to an optical spectrum analyzer (OSA, AQ6373B, Yokogawa) and the other port was connected to another 90/10 coupler for a  photodetector and a power meter. A digital acquisition card connected to a computer was used to monitor the transmission spectra in real-time.

\begin{figure*} [ht]
\centering
      \includegraphics[width=1.0\textwidth]{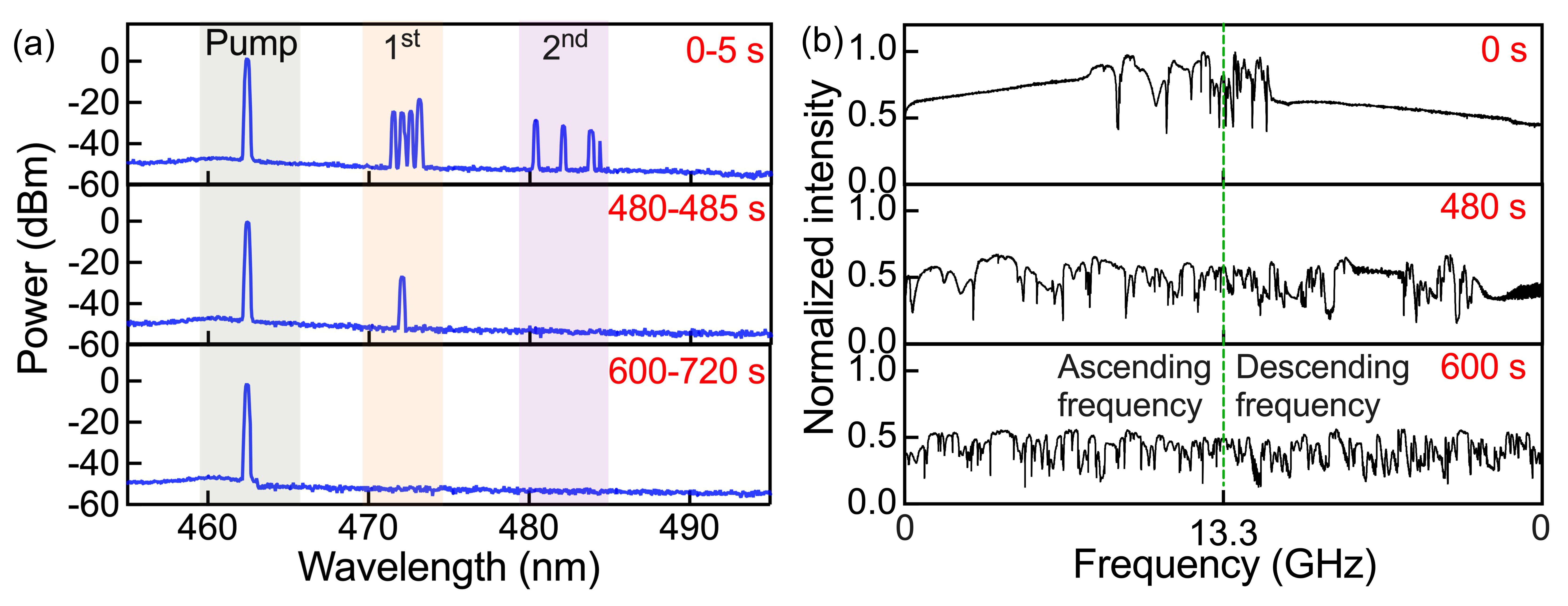}
  \caption{{Blue-band Raman lasing} (a) Recorded stimulated Raman scattering spectra with max hold mode for 5 seconds at different times. (b) Corresponding transmission spectra through the nanofiber at 0 s, 480 s, and 600 s.
  \label{Raman_time}}
\end{figure*}

For the measurements, a silica microsphere  with a diameter of $\sim$100 $\mu$m was fabricated from 460HP SMF using a CO$_2$ laser \cite{microsphere_fabrication}. The Q-factor was $1.8\times10^7$, measured at a low pump power (5 $\mu$W). To generate the SRS, the frequency of the 462~nm pump laser was scanned back and forth over 13.3 GHz. Since the laser frequency was scanning, the Raman spectra were continuously acquired in 5 second intervals using the maximum hold mode of the OSA to ensure most of the Raman peaks were captured, and SRS was observed at a pump power of 12 mW. As shown in Fig. \ref{Raman_time}(a), two orders of SRS were visible during the first 5 s; however, the second order  disappeared after 480 s and no SRS could be observed after 600 s. 

The evolution of the SRS spectra can be understood by monitoring the corresponding WGM spectra, shown in Fig. \ref{Raman_time}(b). the spectrum was dominated by thermally broadened modes that covered nearly the entire range of the laser scan; strong thermal broadening is associated with high Q modes. As time progressed, the high Q modes degraded and their thermal broadening decreased, allowing more high Q modes to enter the scan range. Ultimately, however, the losses began to overcome all the modes - after 600 s  thermal broadening was negligible and  SRS could no longer be observed. This phenomenon should not be  attributed to the degradation of the tapered fiber itself. We deduce that it is due to a photodarkening-induced reduction of the microcavity Q-factor that led to the annihilation of the SRS signal. Simultaneously, the transmission through the fiber decreased to 40$\%$ due to a build-up of scatterers which in turn reduced the thermal broadening. However, on monitoring  mode shifting of low Q modes and  thermal broadening of  high Q modes, the following was clear:  While there was minimal thermal shift of the spectrum,  thermal broadening of the high Q modes changed dramatically and dynamically, indicating that  the Q-factor of the high Q modes was strongly affected.


Critically, once a microsphere was pumped with blue laser light, it was degraded to such a point that it could no longer support SRS. This was verified by replacing the nanofiber with a newly fabricated one (with the same dimensions) and then pumping the degraded microsphere with the same power (12 mW). We found that SRS was no longer achievable even for different coupling positions and laser detunings.


\begin{figure*} [ht]
\centering
      \includegraphics[width=0.5\textwidth]{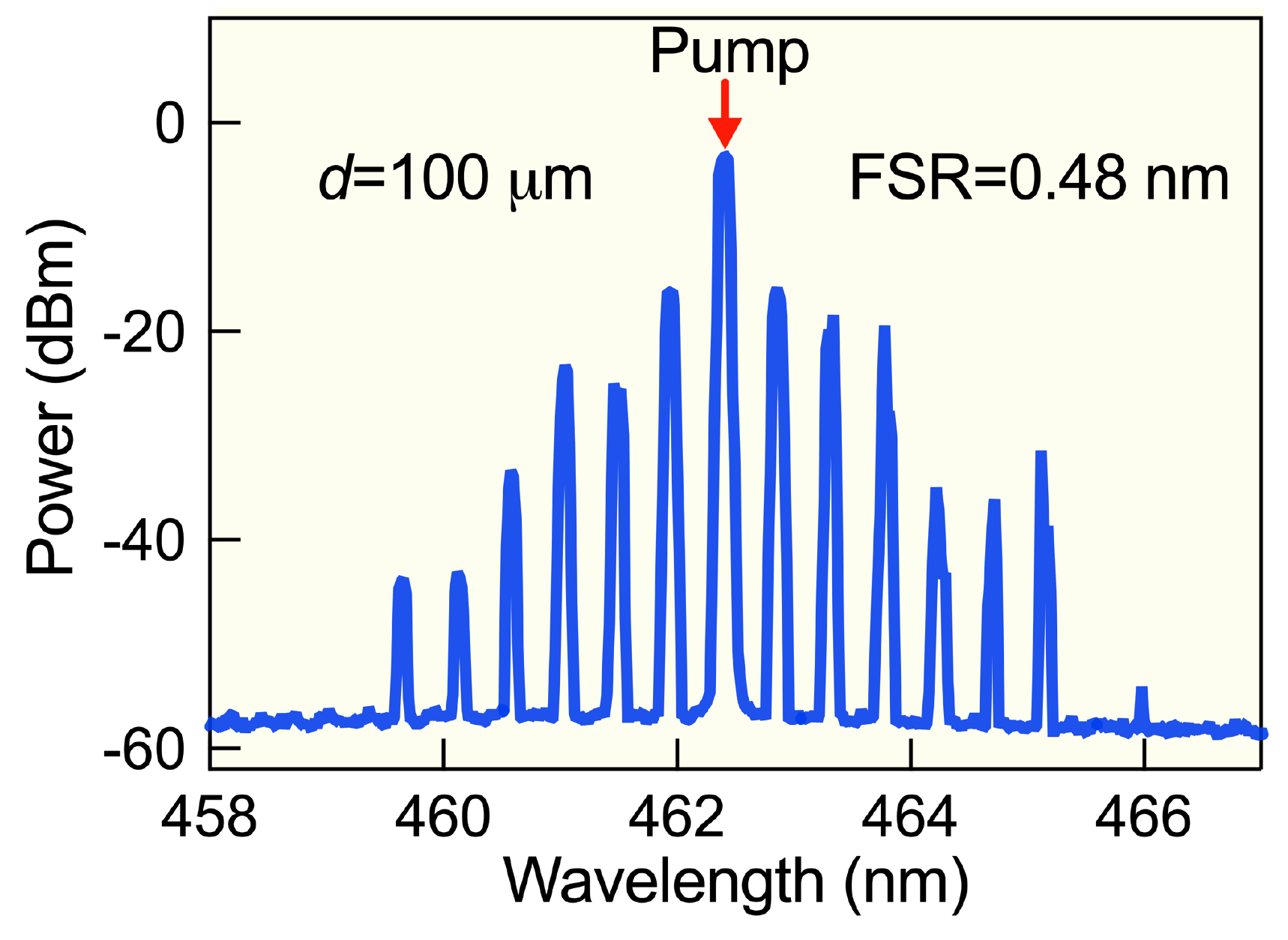}
  \caption{The observed frequency comb around 462 nm.  \textit{d} is the diameter of the silica microsphere and the FSR is about 0.48 nm. 
  \label{holdcomb} }
\end{figure*} 

Apart from SRS, a Kerr nonlinearity based frequency comb in the blue-band was also studied. A new microsphere was made and coupled to a nanofiber by scanning the frequency of the pump laser and adjusting the polarization state at the fiber waist. A Kerr frequency comb with 13 teeth was excited for an input power of $\sim$30 mW, see Fig. \ref{holdcomb}. This is the first time, to the best of our knowledge, a Kerr frequency comb was created in the blue-band region in an SiO$_2$ WGR. The measured free spectral range (FSR) was 0.48 nm, which agrees well with the calculated FSR of 0.46 nm. It is worth noting that it is usually challenging to achieve blue-band hyperparametric oscillation and a Kerr frequency comb in SiO$_2$ resonators because the group velocity dispersion (GVD) of silica is in the normal regime; however, the local dispersion around the pump can be easily modified to be anomalous with the assistance of mode coupling between different transverse modes, which usually exist in microsphere resonators \cite{liu2014investigation,savchenkov2012kerr,lobanov2015frequency}. However, due to the strong photodarkening in the blue-band, the observed Kerr frequency combs were difficult to maintain for a long period of time. The unstable and dynamic behavior of the high Q modes under blue light excitation produced an unsustainable comb, which only lasted for a few seconds. This indicates that photodarkening is an obstacle for directly extending   silica micro/nanophotonics from the IR into the visible band, even for ultralow-threshold nonlinear optics applications. Next, we developed a method that allowed us to stabilize the nanofiber transmission and the WGMs of the microsphere resonator by \textit{in-situ} heating to achieve lasting SRS and hyper-parametric oscillation. 



\subsection{Thermal Bleaching of Photodarkening in Silica Nanofibers}

As reported in the literature, the effect of photodarkening can be reduced by thermal and photobleaching \cite{thermalbleachingcolorcenter, Liu:20,wavelength,thermal-bleaching,thermal-bleaching2,1995photoinduced}. The color centers are strongly influenced by temperature over a wide range. For example, the concentration of the defects is strongly dependent on temperature over the range 10K to above 600K \cite{thermal-bleaching2, SiO2glass}, probably due to diffusion and reactions with mobile hydrogen in the glass\cite{F2laser}. The temperature-dependent mobility of atomic oxygen and its role in the formation of various color centers and concentration is also noted \cite{SiO2glass}. 
we found that apparent complete bleaching of photodarkening in nanofibers could be achieved by a small increase in the ambient temperature while the fiber was optically pumped. Even weak optical pumping is expected to increase the temperature at the waist of a nanofiber in air \cite{taperedfibercavity}. To explore thermal bleaching in nanofibers and WGRs, the temperature around the nanofibers was changed using a metal ceramic heater. A thermocouple was placed at a distance of $\sim$100 $\mu$m from the waist of the nanofibers to record the ambient temperature.

\begin{figure} [ht]
\centering
      \includegraphics[width=1.0\textwidth]{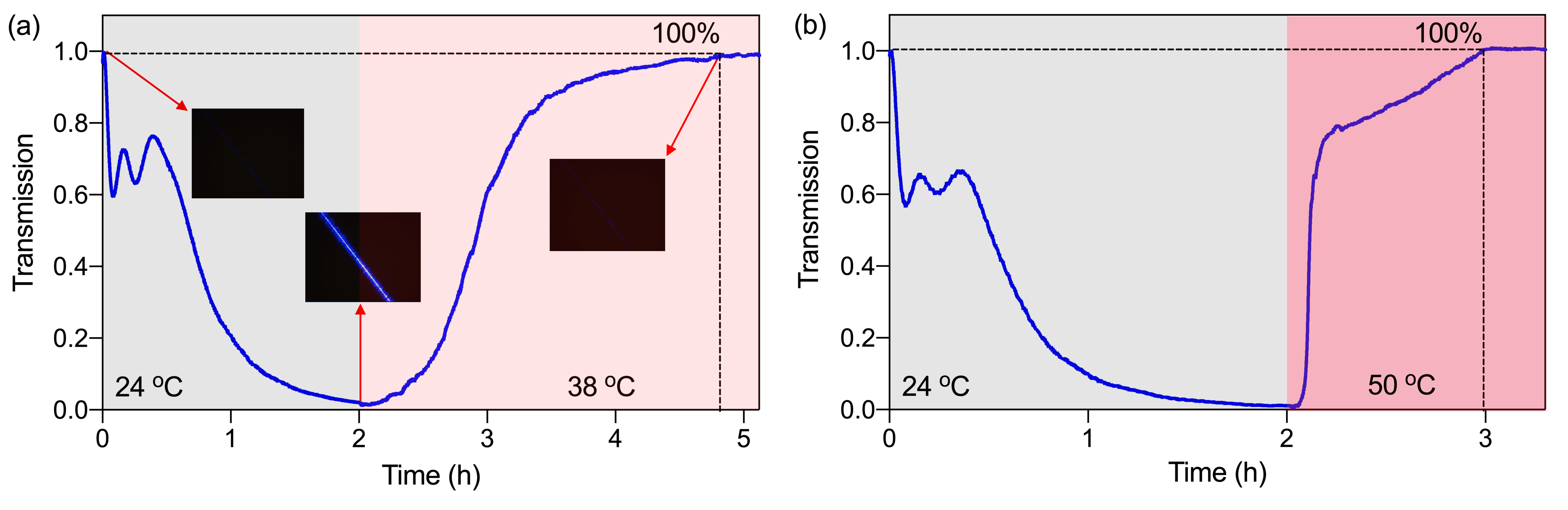}
  \caption{{Thermal bleaching of a nanofiber after  photodarkening}. The transmission through the nanofiber for different temperatures.  (a) 24-38$^\circ$C. Inset: images of the waist of the nanofiber at  0/120/290 minutes.  The exposure time of the CCD camera was set to the lowest value at the beginning. (b) 24-50$^\circ$C. 
  \label{thermalbleaching} }
\end{figure} 

Figure \ref{thermalbleaching}(a) and (b) show the photodarkening and thermal bleaching processes at different temperatures for an input laser power of 7 mW at 462 nm. The heater was turned off for the first 2 hours and the ambient temperature was 24$^\circ$C.  During this time, the transmission decreased, consistent with the  result in Fig. \ref{fig.1}(d). After 2 hours, when the transmission was reduced to only a few percent, the heater was turned on and the temperature was stabilized at 38$^\circ$C and 50$^\circ$C for two different nanofibers. We observed that the photodarkening could be completely bleached with recovery times of ~167 minutes for 38$^\circ$C and ~58 minutes for 50$^\circ$C. Hence, photodarkening can be overcome in an environment close to room temperature. It is worth noting that one does not need to wait until the appearance of photodarkening to perform  thermal bleaching, i.e., the heater can be turned on immediately to prevent photodarkening from occurring. 

\subsection{Stabilization of Blue-band Nonlinear Optics using Thermal Bleaching}
\begin{figure} [ht]
\centering
      \includegraphics[width=0.9\textwidth]{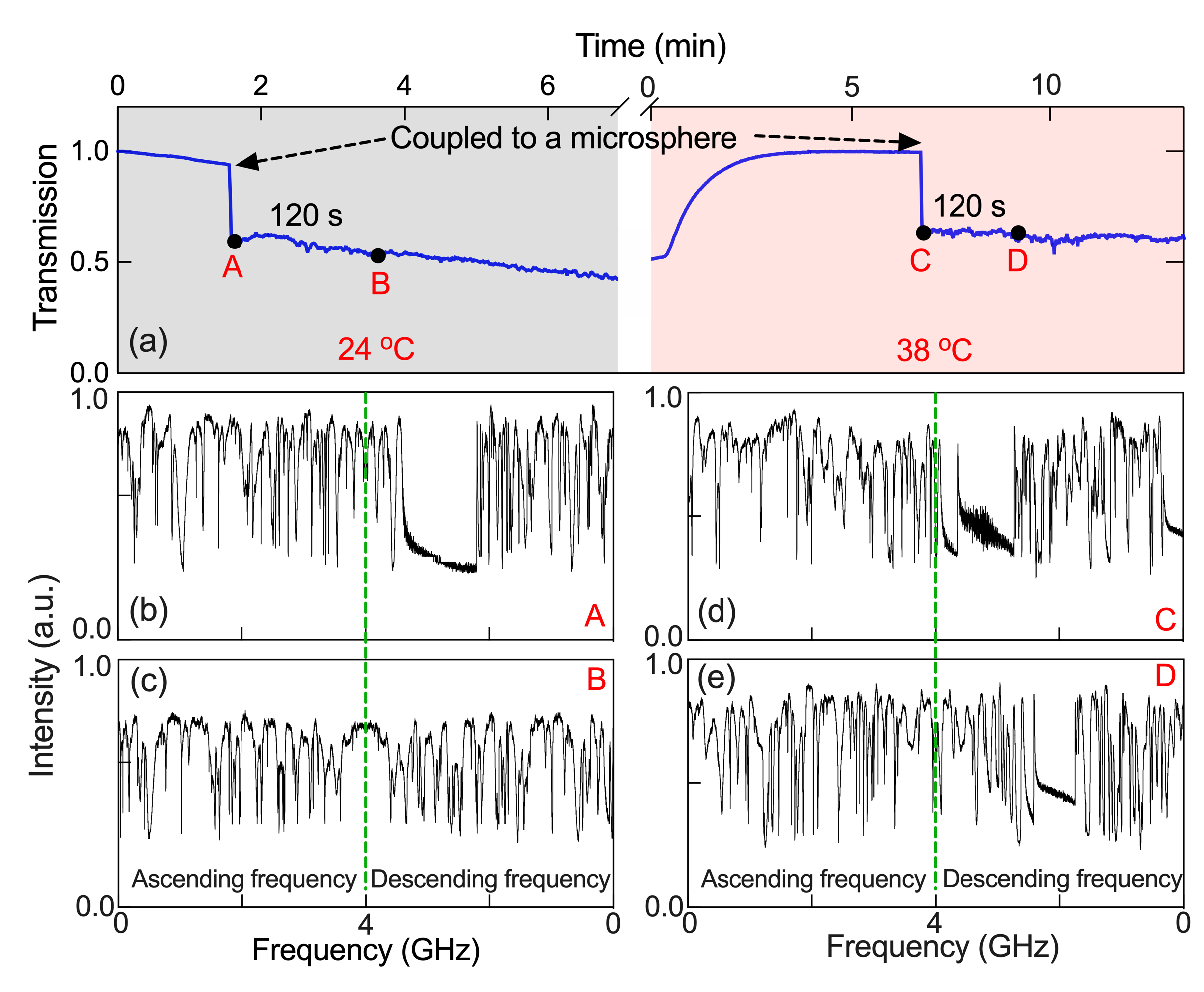}
  \caption{{Operation of a nanofiber-coupled microresonator system at different temperatures}. (a) Recorded transmission through the nanofiber. The time difference between A and B, C and D is 120 s. The corresponding WGM spectra at (b) point A, (c) point B, (d) point C, and (e) point D.
  \label{WGM}}
\end{figure}

Next, we confirmed the role of thermal bleaching in a nanofiber-coupled microresonator system and used the above method to prevent excessive photodarkening from disturbing the nonlinear processes in the cavity. The transmission through the nanofiber and the WGMs spectra were recorded simultaneously, see Fig. \ref{WGM}. The nanofiber transmission was recorded over several minutes at 24$^\circ$C with an input power of 1 mW. It is clear that the transmission through the nanofiber reduced over time, at a rate of  -3.94$\%/$min. At time, A, the microsphere was placed into contact with the nanofiber waist (see the sharp drop in transmission in Fig. \ref{WGM}(a)), and the laser frequency was scanned over 4 GHz at a rate of 10 Hz. Note, at this point, the modes exhibited strong thermal broadening (see Fig. \ref{WGM}(b)), indicating the presence of high Q modes. 120 s later, at time B, the thermal-broadened high Q mode was not present (see Fig. \ref{WGM}(c)).


\begin{figure} [ht]
\centering
      \includegraphics[width=1.0\textwidth]{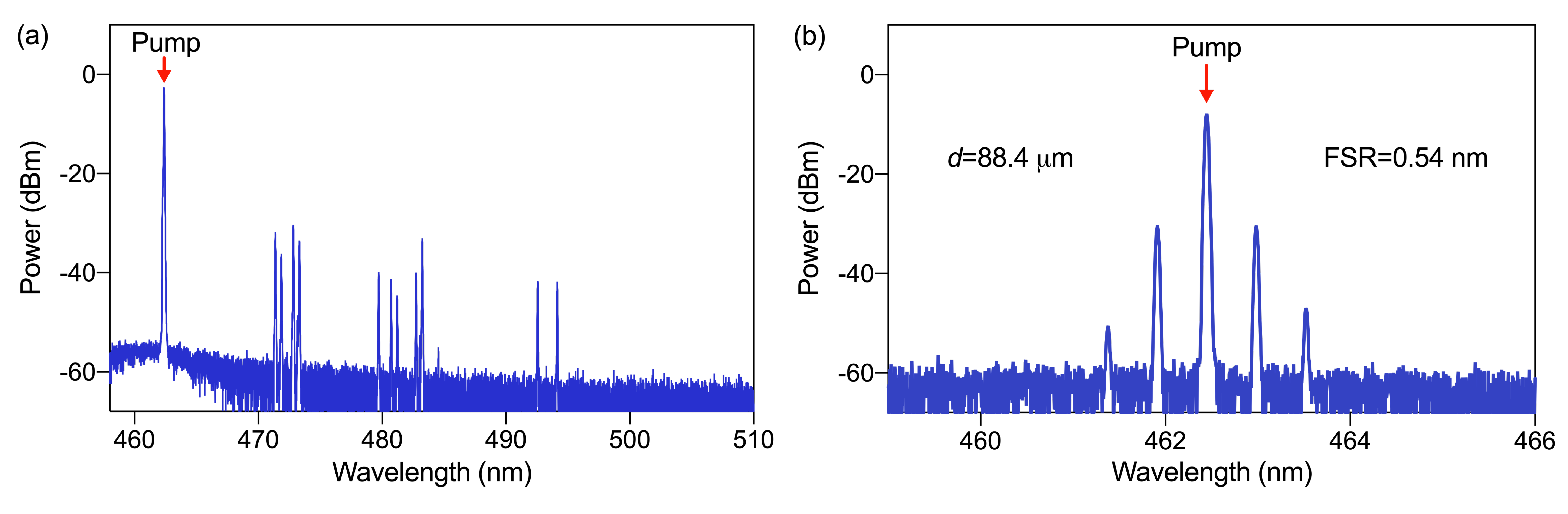}
  \caption{{Generation of (a) stimulated Raman scattering and (b) hyper-parametric oscillation in the blue-band in a silica microsphere. \textit{d} is the diameter of the silica microsphere and the FSR is about 0.54 nm.}  
  \label{fig:figure}}
\end{figure}

The experiment was repeated at 38$^\circ$C with the same nanofiber and microsphere. First,  photodarkening in the nanofiber was thermally bleached before making contact with the microsphere. As the temperature was raised from 24$^\circ$C to 38$^\circ$C, the transmission increased exponentially from 50 to nearly 100 $\%$ in a matter of minutes. Once the transmission was relatively stable, the microsphere was placed in contact with the nanofiber waist, see point C in Fig. \ref{WGM}(a). After the microsphere was coupled, the transmission remained constant (slope -0.58$\%/$min), in contrast to the previous case at 24$^\circ$C. The spectrum in Fig. \ref{WGM}(d) also shows high Q modes; however, unlike for the lower temperature case, these high Q modes persisted, as observed at point D in Fig. \ref{WGM}(e). The stabilised high Q modes and transmission, along with the absence of photodarkening, provided the right conditions for the SRS and hyper-parametric oscillation to occur. When the laser was manually tuned and thermally locked into near-resonance with one of the high Q modes, both stable SRS and hyper-parametric oscillation were achieved, as shown in Fig. \ref{fig:figure}.  Up to three orders of cascaded SRS were realized at a pump power of 13 mW, see Fig. \ref{fig:figure}(a). The spacing between the center of the 1st order SRS band and the pump light is $\sim$10 nm ($\sim$13.8 THz) and this is consistent with the Raman shift of bulk SiO$_2$ glass \cite{kasumie}. In Fig. \ref{fig:figure}(b), hyper-parametric oscillation was achieved at a pump power of 15.2 mW. The measured free spectral range (FSR) was 0.54 nm, which agrees with the calculated FSR of 0.53 nm. 
In both cases, the thermal locking and transmission remained stable allowing for SRS and hyper-parametric oscillation to be maintained for several minutes without any intervention. This is in stark contrast to the situation where photodarkening was not controlled by \textit{in-situ} thermal bleaching. The above tests demonstrated the sensitivity of micro and nanophotonic devices to thermal bleaching and the effectiveness of our method which provides the basic conditions for nonlinear research in silica resonators.

\section{Discussion}

There are variety of mechanisms for the generation of color centers that lead to photodarkening at 462 nm.  Generally, it is  considered to be due to intrinsic stable defects in the form of 'E' centers, oxygen deficiency centers (ODCs), oxygen (excess) interstitial centers, or peroxy radicals \cite{1991optical,Ge1,silicacolorcenters}. These defects have numerous absorption bands; however, of particular interest to optical pumping at 462~nm, are the 'E' center absorption bands at 210 nm and 250 nm, and the 245 nm absorption band of the divalent ODC(II) \cite{silicacolorcenters}, which has a photoluminescence band at 460 nm. The peroxy radical has two adsorptions of interest at 620 nm and 258 nm \cite{1991optical,Ge1, Ultrafastlaser}. Another common color center in pure silica is a partially bound oxygen atom with one free electron, that is the non-bridging oxygen hole center (NBOHC) \cite{stone}. The NBOHC has two absorption bands at 620 nm and 258 nm. This defect often forms as a result of optically induced breaking of a bond in a stressed multiple member Si-O ring \cite{stone}. In Ge-doped silica there are GE1 and GE2 centers with absorption bands around 281 nm and 213 nm, respectively \cite{Ge1}. It is also noted that dangling bonds on the surface of the silica play a role in some laser-induced color centers \cite{silicacolorcenters}. 
In addition to the stable defects, there are unstable self-trapped excitons (STE), which occur when an electron is directly transported to the conduction band, leaving a hole in the valence band. This happens after laser excitation by a multiphoton absorption process. Generation of excitons leads to the creation of additional levels below the conduction band, changing the optical properties of the glass \cite{Ultrafastlaser}. This is a strong candidate for loss at 462 nm since it has absorption peaks at 460 nm; however, it has been noted that the bulk silica absorption can shift to longer wavelengths when the glass is drawn into a fiber \cite{Ge1}(and reference 21 therein). Apart from drawing-induced changes, the defects are thermally sensitive and can migrate and recombine when heated or cooled \cite{SiO2glass}. For example, the movement/displacement of the atomic oxygen from STE sites results in the formation of the E centers and NBOHCs. The unexpected sensitivity of the photodarkening to small changes in temperature observed in our result is not fully explored and beyond the scope of this paper. However, it would seem that nano- and microphotonic SiO$_2$ devices operate in a region where even a few color centres restricts performance. If we now consider that the mobility of defect centers is dependent on temperature and the color center population affects the refractive index \cite{1995photoinduced}, it seems likely that such ultrasensitive devices are subject to the dynamics of these defects.

\section{Conclusion}

In conclusion, blue-band nonlinear optics effects in a nanofiber-coupled silica microresonator were investigated experimentally with hyper parametric oscillation and SRS being demonstrated for the first time in an SiO$_2$ microresonator at 462 nm. Due to the high optical intensity in nanofibers and microresonators, photodarkening is unavoidable and occurs at very low pump powers around 50 $\mu$W. As a result, the fiber transmission, Raman, and Kerr frequency comb signals cannot be maintained for reasonable pump powers. These results imply that photodarkening hinders the advancement of nonlinear photonics at short wavelengths in silica-based devices. Crucially, the photodarkening was demonstrated to be sensitive to small changes in ambient temperature.  Taking advantage of this the \emph{in situ} thermal bleaching proved an effective method for mitigating photodarkening losses, thereby enabling us to study blue-band nonlinear effects in silica resonators. Up to three orders of cascaded SRS and hyper-parametric oscillation were achieved in the presence of a slightly elevated ambient temperature during optical pumping. In addition, the results presented herein also illustrate that the tapered optical fiber could be used as a tool to observe, \emph{in situ}, the dynamics of color center formation. The one dimensional, ultrathin fiber provides us with an easy technique for imaging the process and could facilitate the study of color centers in different materials using sub mW pump powers.

\medskip
\textbf{Supporting Information} \par 
Supporting Information is available from the Wiley Online Library or from the author.

\medskip
\textbf{Acknowledgements} \par 
K. Tian and J. Yu contributed equally to this work. The authors acknowledge S. Kasumie and J.-B. Cl\'ement for early contributions to this work. This research was funded by Okinawa Institute of Science and Technology Graduate University (OIST), China Scholarship Council (201906680), and Harbin Engineering University Scholarship Fund.

\medskip

%

\bibliographystyle{MSP}
\bibliography{report}

\end{document}